\documentclass[12pt]{article}
\pdfoutput=1
\usepackage{graphicx}
\usepackage{amsmath, amssymb}
\usepackage{slashed}
\usepackage{xcolor}    
\usepackage{array,arydshln}
\usepackage{cite}
\usepackage[italicdiff]{physics}
\usepackage[pdftex,
  colorlinks=true,
citecolor=green!60!blue]{hyperref}
  
\begin{document}

\begin{titlepage}

\def\thefootnote{\fnsymbol{footnote}}

\begin{center}

\hfill KANAZAWA-22-02\\
\hfill December, 2022

\vspace{0.5cm}
{\Large\bf Topology-insensitive axion mass in magnetic topological insulators}

\vspace{1cm}
{\large Koji Ishiwata} 

\vspace{1cm}

{\it {Institute for Theoretical Physics,
    Kanazawa University, \\
    Kanazawa 920-1192, Japan}}

\vspace{0.2cm}







\abstract{We study the axion in three-dimensional topological
  insulators with magnetic impurities under finite temperature. We
  find a stable antiferromagnetic ground state and ferromagnetic
  metastable state. In both magnetic states, the mass of the axion is
  found to be up to eV scale and it approaches zero near the phase
  boundary of the magnetic state. This result applies to both normal
  and topological insulator phases, i.e., the axion mass is
  insensitive to the topological states, and it will have a direct
  impact on the targeted mass range of particle axion dark matter in
  future experiments.}

\end{center}
\end{titlepage}

\renewcommand{\theequation}{\thesection.\arabic{equation}}
\renewcommand{\thepage}{\arabic{page}}
\setcounter{page}{1}
\renewcommand{\thefootnote}{\#\arabic{footnote}}
\setcounter{footnote}{0}

\section{Introduction}
\label{sec:intro}

Axion has drawn attention in interdisciplinary fields of particle
physics, cosmology, and condensed matter physics. In condensed matter
physics, a dynamical axion is predicted in the magnetic topological
insulators (TIs)~\cite{Li:2009tca}.  It is a quasi particle that
couples to the electromagnetic fields, which leads to an instability
of the electromagnetic fields. It is predicted that the instability
causes the total reflection of the incident light~\cite{Li:2009tca} or
the conversion of the external electric field to magnetic
field~\cite{Ooguri:2011aa}. On top of that, it was proposed in
Refs.\,\cite{Marsh:2018dlj,Chigusa:2020gfs,Chigusa:2021mci} that the
axion in the magnetic TIs can be a possible excitation signal in the
detection of {\it particle} axion, which is a good candidate for dark
matter of the universe. On the other hand, a static axion or constant
axion is known as the magnetoelectric
effect~\cite{Qi:2008ew,Qi:2008pi,Hehl:2007ut,Dzyaloshinskii,Essin:2008rq,Wu:2016oxw}. To
understand the properties of both the dynamical and static axions,
magnetism plays a crucial role.

In Refs.\,\cite{Sekine:2014xva,Sekine:2015eaa,Schutte-Engel:2021bqm},
the dynamical axion in the antiferromagnetic (AFM) TIs is described
from the partition function given by the path integral.  As a result,
the mass of the dynamical axion in the AFM TIs is estimated to be
about a meV.  This mass range corresponds to the projected mass range
of {\it particle} axion proposed by Ref.\,\cite{Marsh:2018dlj}.  On
the other hand, Ref.\,\cite{Ishiwata:2021qgd} has revisited the axion
mass in the Hubbard model and reformulated the action of the axion
field using the Hubbard-Stratonovich transformation. In the formula,
the effective potential for the axion field is derived in the
topological and normal insulators under the AFM and paramagnetic
states.  Consequently, both the dynamical and static axions are
described consistently and the axion mass is found to be less than
$\order{\rm eV}$. Furthermore, it can be suppressed near the phase
boundary between the AFM and paramagnetic states.  Since the axion
mass in materials directly corresponds to the mass range of {\it
  particle} axion in the proposal of any future axion detection
experiment~\cite{Marsh:2018dlj}, the evaluation of the axion mass in
materials is crucial.

In recent years, magnetically doped bismuth selenide or bismuth
telluride has caught lots of attention. For example, both the AFM and
ferromagnetic (FM) states are predicted in
MnBi$_2$Te$_4$~\cite{Li:2019j,Li:2019h,Li:2019jia,Zhang:2019don,Hao:2019}
or Mn$_2$Bi$_2$Te$_5$~\cite{Li:2020fvr} by the first-principles
calculations. Regarding Mn$_2$Bi$_2$Te$_5$, a rich magnetic
topological state in addition to the AFM/FM states are
predicted~\cite{Li:2020fvr}.\footnote{The dynamical axion was studied in
Ref.\,\cite{Zhang:2019lkh}, and recently Ref.\,\cite{Cao:2021} reported
that Mn$_2$Bi$_2$Te$_5$ is synthesized and its experimental aspects are
studied.} Since such materials are probable candidates for the
detection of the particle axion, it is important to find how to
describe the axion in a variety of magnetic states.

In this work, we formulate the axion in the TIs with magnetic dopants
under finite temperature.  For this purpose, we consider the
three-dimensional (3D) effective TIs model with the interaction term
of electrons with magnetic impurities. In this study we do not specify
the explicit material. The grand potential is calculated from the path
integral under finite temperature, and consequently, the effective
potential for the order parameter of the AFM and FM are
derived. Around the stationary points of the effective potential, the
mass of the dynamical axion is formulated. We will see that the
typical mass scale of axion in the magnetic insulators is eV and it
can be suppressed near the phase boundary, depending on
temperature. This feature is insensitive to the topological states of
insulators. As a check, we will also see that the curvatures with
respect to the order parameters of the AMF and FM phases correspond to
the Van Vleck-type spin susceptibility, which is calculated in the
linear perturbation theory, for a band insulator.  The result implies
that the mass range of the {\it particle} axion which is projected to
be probed~\cite{Marsh:2018dlj} is around eV scale and it depends on
the magnetic state of the insulators. The temperature dependence of
the axion mass might be used to search for low-mass regions.

This paper is organized as follows. In the next section, we give the
Hamiltonian of the model and define the observables including the
order parameters. In Sec.\,\ref{sec:Veff} the effective potential for
the order parameters is derived from the grand potential, which gives
rise to the phase diagram of magnetism in
Sec.\,\ref{sec:phase}. Finally, the axion mass is derived in
Sec.\,\ref{sec:AxionMass}. The conclusion is given in
Sec.\,\ref{sec:conclusion}.

\section{The model}
\label{sec:model}
\setcounter{equation}{0} 

We consider a effective model for 3D topological insulators (TIs). The
basic Hamiltonian is~\cite{Zhang:2009zzf,Li:2009tca}\footnote{We
change the notation of Hamiltonian from one in
Ref.\,\cite{Ishiwata:2021qgd}.}
\begin{align}
  H^{\rm TI}
  &= \sum_{\vb*{k}}c^\dagger_{\vb*{k}}{\cal H}^{\rm TI}_{\vb*{k}} c_{\vb*{k}}\,,
  \\
  {\cal H}^{\rm TI}_{\vb*{k}} &= (\epsilon_0-\mu){\bf 1}
  +\sum_{a=1}^4d^a\Gamma^a\,,
  \label{eq:H0}
\end{align}
where $c_{\vb*{k}}^\dagger$ and $c_{\vb*{k}}$ are the creation and
annihilation operators of electrons in the wavenumber space and $\mu$ is
the chemical potential, $\vb*{k}$ is the wavenumber, and $\Gamma^a$ are
the Gamma matrices defined in Eq.\,\eqref{eq:Gamma1-4} of
Appendix~\ref{sec:Gammas}.  $\epsilon_0$ is a constant and $d^a$ is
parameterized as
\begin{align}
  (d^1,d^2,d^3,d^4) =
  (A_2 \sin k_x \ell_x,\, A_2 \sin k_y\ell_y,\,A_1 \sin k_z\ell_z,\,{\cal M})\,,
\end{align}
where ${\cal M}=M_0-2B_1-4B_2+2B_1 \cos k_z\ell_z+2B_2(\cos k_x\ell_x
+ \cos k_y\ell_y)$.  $M_0<0$ and $M_0>0$ correspond to topological and
normal insulators, respectively.  We consider a cubic lattice in the
later analysis, i.e. $\ell_x=\ell_y=\ell_z\equiv \ell$, for
simplicity.  The Hamiltonian has the time-reversal invariance, which
is one of the features of the TIs, and it describes the Bi$_2$Se$_{3}$
family of materials, including Bi$_2$Te$_{3}$ and
Sb$_2$Te$_{3}$~\cite{Liu:2010}. In the present study, we additionally
assume magnetic dopants, such as Fe, Cr, or Mn, in the material and
introduce the onsite interaction term between the impurity and
electron~\cite{Yu:2010hth}
\begin{align}
  H_J = \sum_{I}^{N_s} \bigl[
    J^A\vb*{S}^A(\vb*{x}_I)\vdot \vb*{s}^A_I
    +J^B\vb*{S}^B(\vb*{x}_I)\vdot \vb*{s}^B_I\bigr]\,,
\end{align}
where $\vb*{S}^A$ ($\vb*{s}^A_I$) and $\vb*{S}^B$ ($\vb*{s}^B_I$) are
the local spins of the impurities (spins of electron) at cite $I$ of
the sublattices $A$ and $B$, respectively. $J^A$ and $J^B$ are the
exchange coupling constants and $N_s$ is the number of the impurities.
In the following discussion, we consider the magnetism in the $z$
direction.  Then the spins of electron are written as
\begin{align}
  s^A_{zI} &= \frac{1}{2}c^\dagger_I (\Gamma^{12}+\Gamma^5)c_I\,, \\
  s^B_{zI} &= \frac{1}{2}c^\dagger_I (\Gamma^{12}-\Gamma^5)c_I\,,
\end{align}
where $c_I$ is the wavefunction of the electron at cite $I$ in the
lattice space, and $\Gamma^{12}$ and $\Gamma^5$ are given in
Appendix~\ref{sec:Gammas}. The similar model, but only with a term
proportional to $\Gamma^{12}$ is considered in
Refs.\,\cite{Rosenberg:2012,Kurebayashi:2014,Wang:2015hhf} in a
different context. In Ref.\,\cite{Wang:2015hhf}, the same terms as
both $\Gamma^{12}$ and $\Gamma^{5}$ are considered. In the literature,
Cr and Mn are doped on the top and the bottom halves of the TI films
in superlattice and the exchange couplings with Cr and Mn are taken to
be opposite each other. Then the mass of the dynamical axion is
estimated to be meV. We will get a different result in
Sec.\,\ref{sec:AxionMass}.

We apply the mean-field approximation (MFA) to $H_J$.  In the MFA,
$H_J$ becomes
\begin{align}
  H_J\approx
  \sum_I^{N_s}&\bigl[ J^A\expval{S^A_z}s_{zI}^A + J^B\expval{S^B_z}s_{zI}^B
    + J^AS^A_z(\vb*{x}_I) \expval{s^A_z} + J^BS^B_z(\vb*{x}_I) \expval{s^B_z}
    \bigr]
  \nonumber \\
  &-N_s (J^A\expval{S^A_z}\expval{s^A_z} + J^B\expval{S^B_z}\expval{s^B_z})\,.
\end{align}
Introducing
\begin{align}
  M^A = x \expval{S^A_z}\,, ~~ M^B = x \expval{S^B_z}\,,
  \\
  m^A = \expval{s^A_z}\,,~~ m^B = \expval{s^B_z}\,,
\end{align}
where $x=N_s/N$ and $N$ is the number of cite, we get
\begin{align}
  H_J\approx \sum_i^N &\left[
    J^AM^A s_{zi}^A + J^BM^B s_{zi}^B\right]
  +\sum_I^{N_s}\left[
    J^A m^A S^A_{z}(\vb*{x}_I) + J^B m^B S^B_{z}(\vb*{x}_I)\right]
  \nonumber \\
  &-N(J^AM^Am^A+J^BM^Bm^B)\,.
\end{align}
As a result, the total Hamiltonian is linearized as\footnote{Although
$M^A$, $M^B$, $m^A$ and $m^B$ themselves should be interpreted as the
mean-field values (or vacuum expectation values), we take them as
spurious fields to give the effective potential. See the later
discussion. }
\begin{align}
H^{\rm TI} + H_J \approx H_e + H_S+H_R\,,
\end{align}
where $H_e$ and $H_S$ are the Hamiltonians of the electrons and
the local spin defined by
\begin{align}
  H_e &= H^{\rm TI} + \sum_i^N\left[
    J^AM^A s_{zi}^A + J^BM^B s_{zi}^B \right]\,,
  \label{eq:H_e}
  \\
  H_S &= \sum_I^{N_s}\left[
    J^A m^A S^A_{z}(\vb*{x}_I) + J^B m^B S^B_{z}(\vb*{x}_I)\right]\,,
  \label{eq:H_S}
  \\
  H_R &= -N(J^AM^Am^A+J^BM^Bm^B)\,.
   \label{eq:H_R}
\end{align}

For later analysis, it is convenient to write down the Hamiltonian
by using the following variables
\begin{align}
  m_t &= m^A + m^B\,,
  \label{eq:mt}
  \\
  m_r &= m^A - m^B\,,
  \label{eq:mr}
  \\
  M_f &= \frac{1}{2}(J^AM^A+J^BM^B)\,,
  \label{eq:Mf}
  \\
  M_5 &= \frac{1}{2}(J^AM^A-J^BM^B)\,.
  \label{eq:M5}
\end{align}
$M_f$ and $M_5$ plays the order parameters of the FM and AFM,
respectively.  In terms of $M_f$ and $M_5$ the Hamiltonian of
the electrons is given by 
\begin{align}
  H_{e} = \sum_{\vb*{k}}c^\dagger_{\vb*{k}}{\cal H}_{e\vb*{k}}c_{\vb*{k}}\,,
\end{align}
where\footnote{$M_5$ corresponds to $\phi$ in
Ref.\,\cite{Ishiwata:2021qgd}.}
\begin{align}
  {\cal H}_{e\vb*{k}} &= {\cal H}^{\rm TI}_{\vb*{k}} + {\cal H}^m_{\vb*{k}}\,,
  \\
  {\cal H}^m_{\vb*{k}} &= M_f \Gamma^{12}  + M_5 \Gamma^5\,.
\end{align}
We note that the two terms proportional to $\Gamma^{12}$ and
$\Gamma^5$ appear in the Hamiltonian for the electrons. Those terms
describe the magnetism of the materials and they are consistent with
the symmetry of the crystal structure of the materials, such as
Bi$_2$Se$_3$ and Bi$_2$Te$_3$~\cite{Liu:2010}. Diagonalizing ${\cal
  H}_{e\vb*{k}}$ gives four energy bands. They are given by
$E_{j\vb*{k}}=\epsilon_0-\mu \pm e_{j\vb*{k}}$ ($j=1,2$), where
\begin{align}
  e_{1\vb*{k}}&=\sqrt{d_0^2+M_f^2+M_5^2+2M_f\sqrt{d_s^2+M_5^2}}\,,
  \label{eq:e^1}
  \\
  e_{2\vb*{k}}&=\sqrt{d_0^2+M_f^2+M_5^2-2M_f\sqrt{d_s^2+M_5^2}}\,,
  \label{eq:e^2}
\end{align}
where $d_0\equiv \sqrt{\sum_{a=1}^4d^ad^a}$ and $d_s \equiv
\sqrt{(d^3)^2+(d^4)^2}$.

In the following discussion, we consider the half-filling case since
we are interested in the insulator in the bulk.  In addition we assume
that the temperature is sufficiently smaller than the energy scale of
the electron. This is a good approximation since we consider
temperature up to $\order{10^2\,{\rm K}}$.  Then the chemical
potential should be chosen as $\mu\simeq \epsilon_0$, and the relevant
energy bands in the following discussion are going to be
$-e_{1\vb*{k}}$ and $-e_{2\vb*{k}}$.

\section{The effective potential from the grand potential}
\label{sec:Veff}
\setcounter{equation}{0} 

While the mean-field values for each variable can be derived from the
Hamiltonian, the grand potential is useful to derive the effective
action for the order parameters $M_f$ and $M_5$.  The grand potential
is given by\footnote{Since we consider the half-filling case, the
grand potential corresponds to the Helmholtz free energy. }
\begin{align}
  \Omega=-\beta^{-1} \ln Z\,,
\end{align}
where $\beta=1/T$ is the inverse temperature and $Z$ is the partition
function given by
\begin{align}
  Z=\int
  {\cal D}c^\dagger{\cal D}c{\cal D}M
  ~e^{-S_E}\,.
\end{align}
Here $c$ is the wavefunction of the electrons, $M$ represents $S^A_z$ and
$S^B_z$, and $S_E$ is the action of the system in the Euclidean
space.\footnote{In the derivation of the kinetic term of the dynamical
axion, we promote $M_5$ to a dynamical field. See the later discussion and
Appendix~\ref{sec:stiffness}.} The Hamiltonian of the electrons and local
spins are linearized under the MFA, as seen in the previous
section. Then the Euclidean action is given by $S_E =
S_{e}+S_{S}+S_{R}$ where
\begin{align}
  S_{e}&=\int^\beta_0d\tau
  \sum_i^N c_i^\dagger [\partial_\tau+{\cal H}_e]c_i\,,
  \\
  S_{S } & = \int^\beta_0d\tau H_S\,,
  \\
  S_{R } & = \int^\beta_0d\tau H_R=\beta H_R\,.
\end{align}
Consequently, the grand potential is obtained as
\begin{align}
  \Omega = \Omega_e + \Omega_S + H_R\,,
\end{align}
where $\Omega_e$ is the grand potential for the electrons and
$\Omega_S$ is the one for the local spins given by
\begin{align}
  \Omega_S &= -\beta^{-1} N_s\left[
    \ln \frac{\sinh (S+1/2)\beta J^Am^A}{\sinh \beta J^Am^A/2}
    +(A\to B)
    \right]\,.
\end{align}
Here $S$ is the absolute value of the local spin. $\Omega_e$, on the
other hand, is computed as
\begin{align}
  \Omega_e &=-\beta^{-1}\ln e^{-{\cal S}_e}\,,
  \\
  {\cal S}_{e} &= -\ln \det [\partial_\tau+{\cal H}_e]\,.
\end{align}
Here the determinant is obtained by
\begin{align}
  \det[\partial_\tau +{\cal H}_e]
  =\prod_{n} \prod_{j,\vb*{k}}(-i\omega_n+E_{j\vb*{k}})\,,
\end{align}
where $\omega_n=(2n+1)\pi/\beta$ is the Matsubara frequency for
fermions.

From the grand potential, the mean-field (MF) values for $m^A$, $m^B$,
$M^A$, and $M^B$ are obtained as
\begin{align}
  m^A_{\rm MF} &= \frac{1}{N}\pdv{\Omega_e}{J^A M^A}\,,
  \label{eq:mAMF}
  \\
  m^B_{\rm MF} &= \frac{1}{N}\pdv{\Omega_e}{J^B M^B}\,,
  \label{eq:mBMF}
  \\
  M^A_{\rm MF} &= \frac{1}{N}\pdv{\Omega_S}{J^Am^A}\,,
  \label{eq:MAMF}
  \\
  M^B_{\rm MF} &= \frac{1}{N}\pdv{\Omega_S}{J^Bm^B}\,.
  \label{eq:MBMF}
\end{align}
They are also derived from
$\pdv*{\Omega}{M^A}=\pdv*{\Omega}{M^B}=\pdv*{\Omega}{m^A}=\pdv*{\Omega}{m^B}=0$.
In terms of $m_t$, $m_r$, $M_f$ and $M_5$, the MF values are given as
\begin{align}
    m_{t,{\rm MF}}
  &=\frac{1}{N}\pdv{\Omega_e}{M_f}
  \nonumber \\
  &=-\frac{1}{N}\sum_{\vb*{k}} \left[
    \frac{M_f+\sqrt{d_s^2+M_5^2}}{e_{1\vb*{k}}}n_F(E_{1\vb*{k}})
    +\frac{M_f-\sqrt{d_s^2+M_5^2}}{e_{2\vb*{k}}}n_F(E_{2\vb*{k}})\right]\,,
  \label{eq:mtMF}
  \\
  m_{r,{\rm MF}}
  &=\frac{1}{N}\pdv{\Omega_e}{M_5}
  \nonumber \\
  &=-\frac{1}{N}\sum_{\vb*{k}} M_5\left[
    \frac{1+M_f/\sqrt{d_s^2+M_5^2}}{e_{1\vb*{k}}}n_F(E_{1\vb*{k}}) +
    \frac{1-M_f/\sqrt{d_s^2+M_5^2}}{e_{2\vb*{k}}}n_F(E_{2\vb*{k}})\right]\,,
  \label{eq:mrMF}\\
  M_{f,{\rm MF}}
  &= -\frac{1}{2}xS\bigl[J^AB_S(S\beta J^A m^A)+J^BB_S(S\beta J^B m^B)\bigr]\,,
  \label{eq:MfMF}
  \\
  M_{5,{\rm MF}}
  &= -\frac{1}{2}xS\bigl[J^AB_S(S\beta J^A m^A)-J^BB_S(S\beta J^B m^B)\bigr]\,,
  \label{eq:M5MF}
\end{align}
where $n_F(E)=1/(1+e^{\beta E})$ is the Fermi distribution function
and $B_S$ is the Brillouin function.

Since we are interested in the dynamics with respect to $M_f$ and
$M_5$ around possible stationary points, we put the MF values for the
electron spins $m_t$ and $m_r$ and define the effective action for
$M_f$ and $M_5$ as\footnote{In Appendix~\ref{sec:Effaction} we give another
aspect of the definition of the effective action. }
\begin{align}
  \Omega|_{m_t=m_{t,{\rm MF}},\,m_r=m_{r,{\rm MF}}}\,
  \equiv -\beta^{-1}\ln e^{-{\cal S}_{\rm eff}}\,.
  \label{eq:Seff}
\end{align}
Consequently the effective potential for $M_f$ and $M_5$ is given by
\begin{align}
  V_{\rm eff}(M_f,M_5) =\frac{1}{\beta V} {\cal S}_{\rm eff}
  =\frac{1}{V}\Omega|_{m_t=m_{t,{\rm MF}},\,m_r=m_{r,{\rm MF}}}\,,
  \label{eq:Veff}
\end{align}
where $V$ is the volume of the system.  Here we have omitted the
kinetic terms for the fluctuation around the stationary values for
$M_f$ and $M_5$. The derivation of the kinetic term is given in
Appendix~\ref{sec:stiffness}. We will use the effective potential and
the kinetic term to calculate the axion mass in
Sec.\,\ref{sec:AxionMass}.

\section{Magnetic states}
\label{sec:phase}
\setcounter{equation}{0}

\begin{figure}[t]
  \begin{center}
    \includegraphics[scale=0.22]{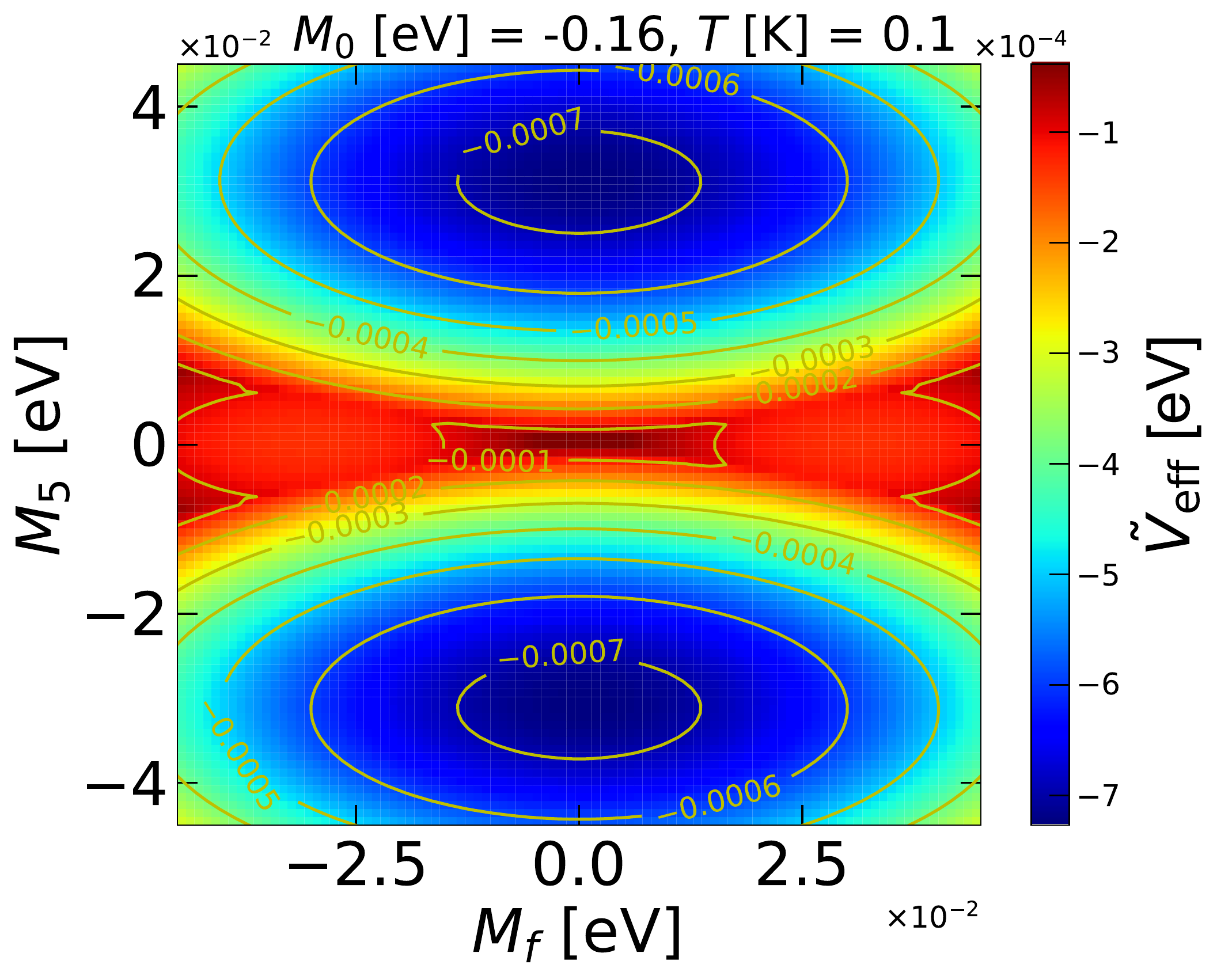}
    \includegraphics[scale=0.22]{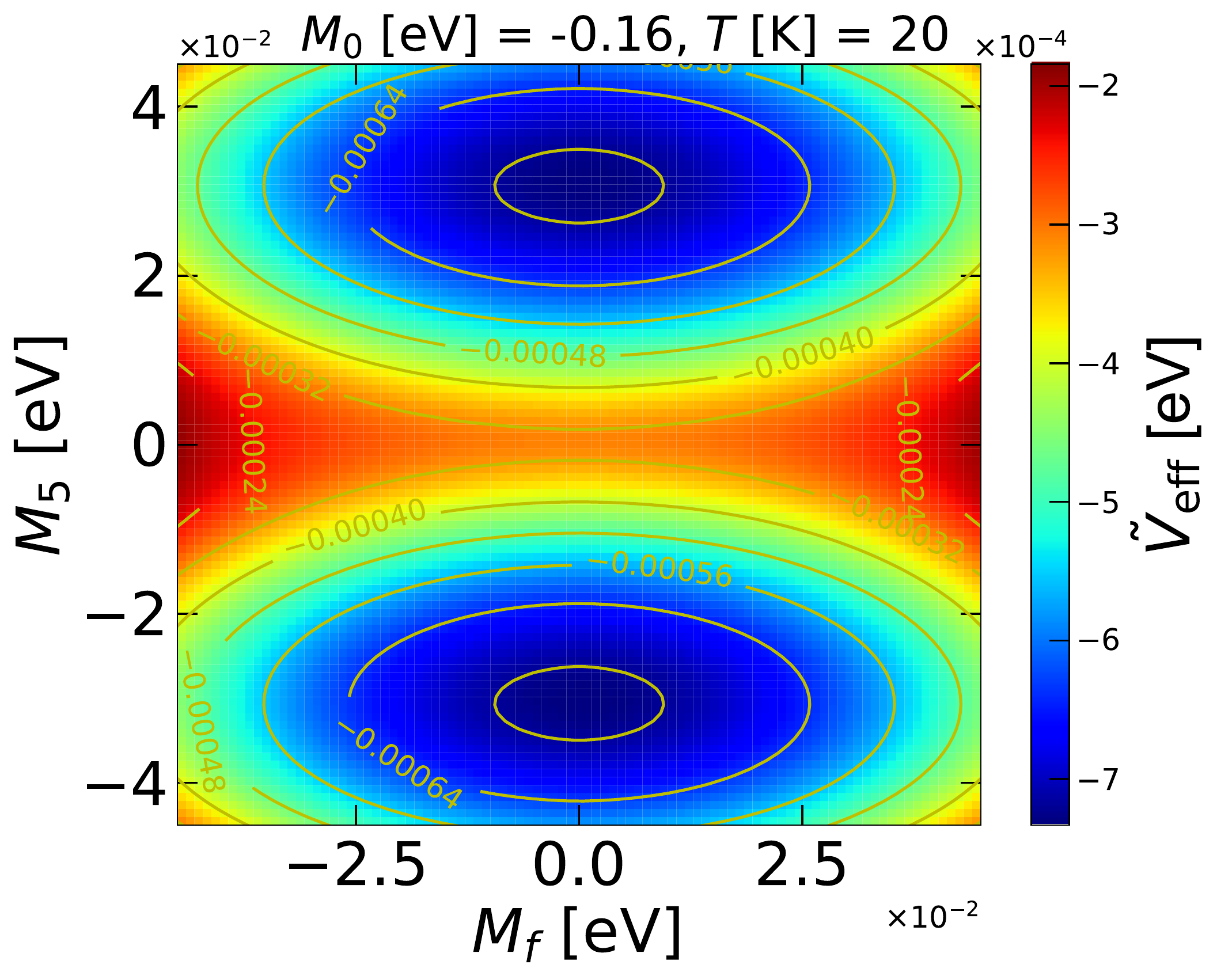}
    \includegraphics[scale=0.22]{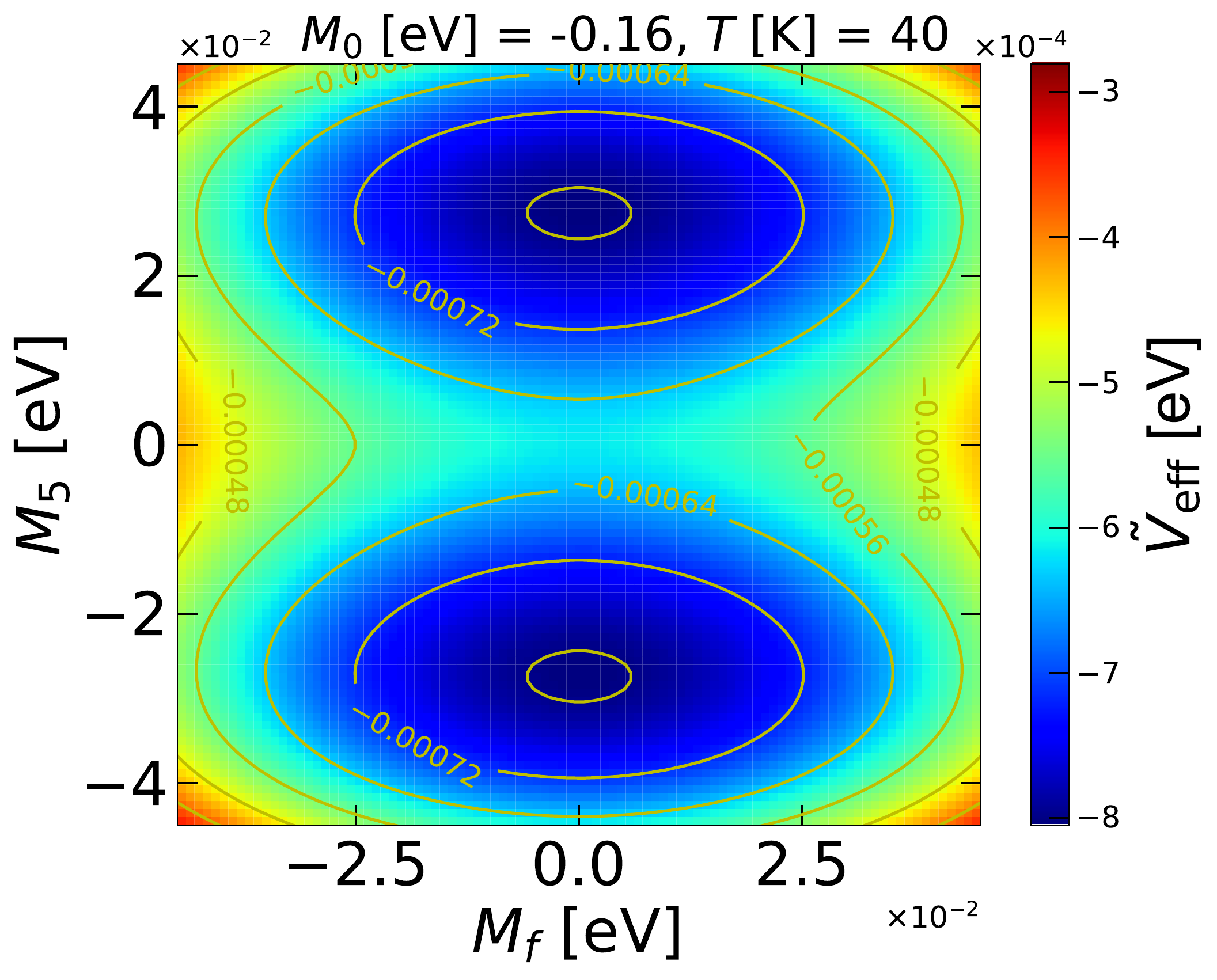}
    \includegraphics[scale=0.22]{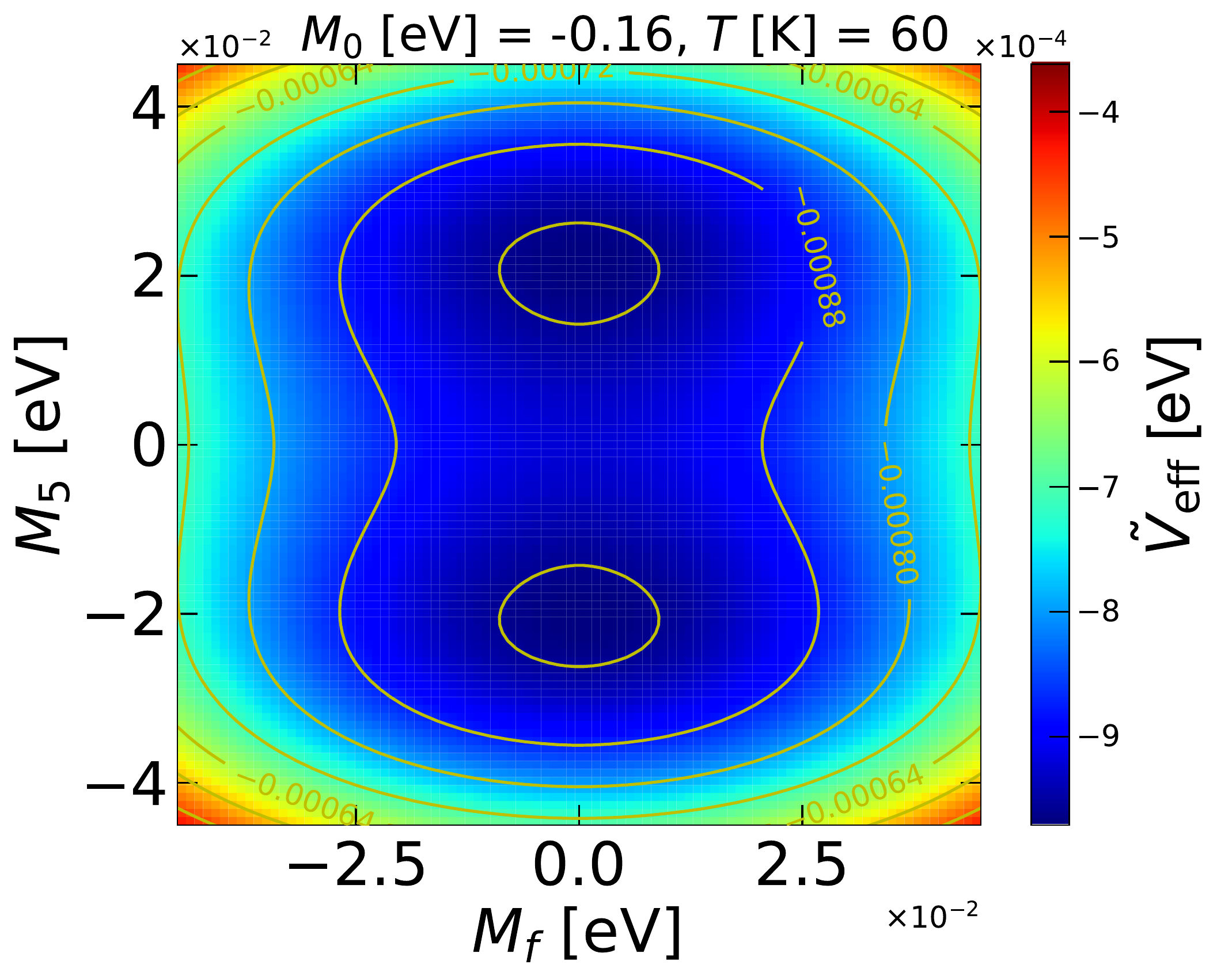}
    \includegraphics[scale=0.22]{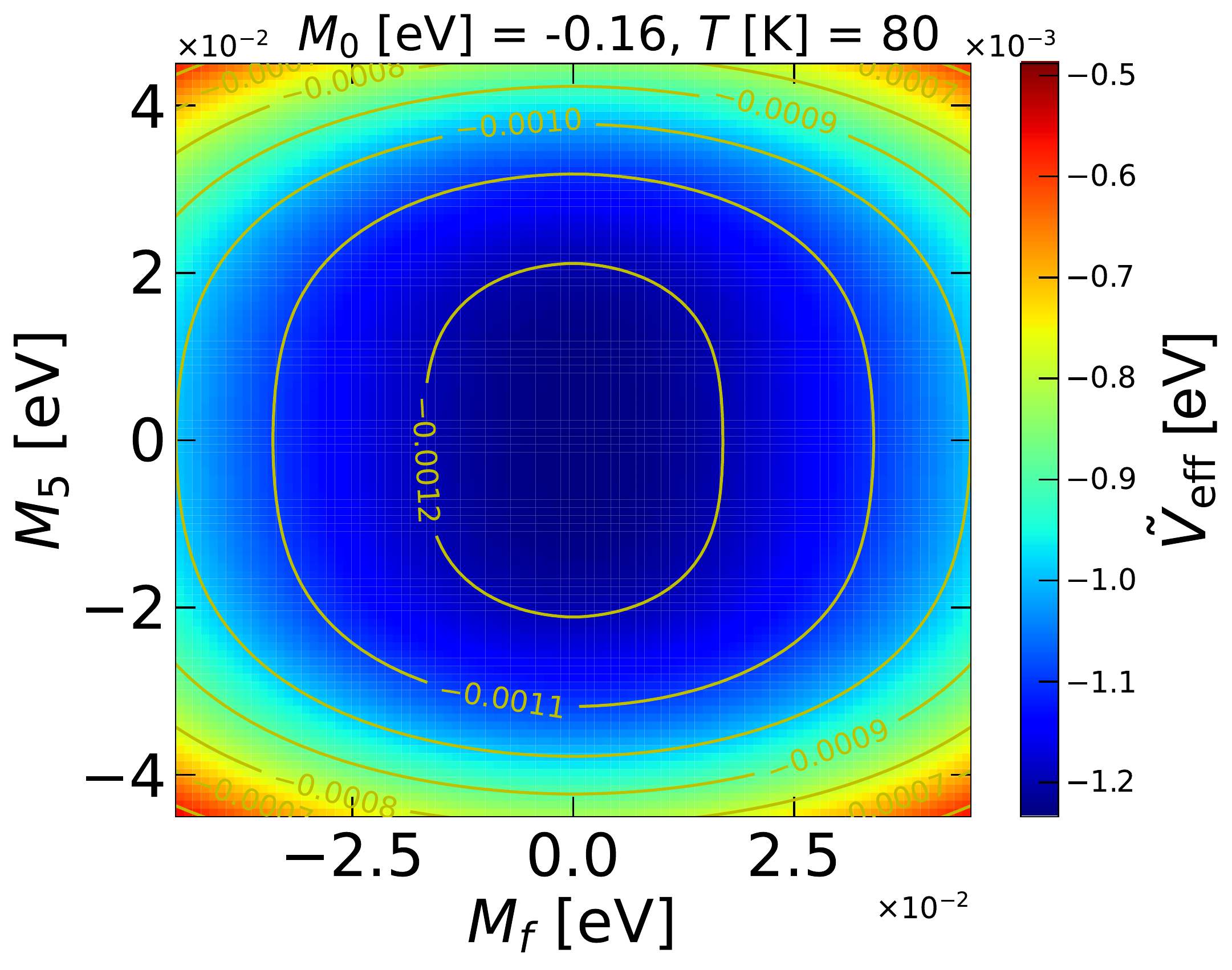}
  \end{center}
  \caption{Color map of the normalized effective potential
    $\tilde{V}_{\rm eff}=[V_{\rm eff}(M_f,M_5) -V_{\rm eff}(0,0)]\ell^3$ on
    $(M_f,M_{5})$ plane. The parameters are $J^A=J^B=0.25$ eV,
    $S=5/2$, $x=0.05$, $A_2=2A_1=0.4$ eV, $B_2=2B_1=-0.4$ eV, and
    $M_0=-0.16$~eV. The temperature is taken to be 0.1, 40, 60 and 80
    K. At each panel, the contours of the potential are shown in solid
    yellow lines.}
  \label{fig:Veff}
\end{figure}

Let us see possible magnetic states, which are determined by the grand
potential or equivalently the effective potential given in
Eq.\,\eqref{eq:Veff}. Fig.\,\ref{fig:Veff} shows the effective
potential on the ($M_f,M_5$) plane for various values of temperature.
In the calculation we take the model parameters as those proposed by
the first-principles
calculation~\cite{Zhang:2009zzf,Fu:2009xfz,Li:2009tca,Rosenberg:2012}
and the values are given in the figure caption. In the figure we plot
the effective potential normalized as $\tilde{V}_{\rm eff}\equiv
[V_{\rm eff}(M_f,M_5) -V_{\rm eff}(0,0)]\ell^3$. For $J^A=J^B>0$ we
found that the global minimum of the potential corresponds to $M_f=0$
and nonzero $M_5=M_{50}$. Here $M_{50}$ is the stationary value for
$M_5$. At the zero-temperature limit, $|M_{50}|= xSJ^A$ is expected
from Eq.\,\eqref{eq:M5MF}, which is consistent with the figure. It is
also clear that the minimum is stable.  This is also checked
analytically, which is shown in Appendix~\ref{sec:ma_zero}. At the
minimum $m_A\simeq -m_B$ is realized, which means that the magnetic
order of the electrons is the AFM. The same is true for $M^A$ and
$M^B$, i.e., $M^A\simeq -M^B$. The AFM disappears for temperature above
the critical temperature $T_{c}^{\rm AFM}$, which is around $80$~K in
the figure. In the region $T>T_c^{\rm AFM}$, the global minimum is at
the origin. Namely, all the MF values are zero and the insulator
becomes the paramagnetic state.

The same result is obtained for $J^A=-J^B$, except that $M^A\simeq
M^B$ is realized. In addition, we find that the effective potential
does not drastically change depending on the sign of $M_0$, i.e., the
topological phase or not. On the other hand, $M_0$ moderately affects
the obserevables, such as critical temperature and axion mass, which
will be discussed below and in Sec.~\ref{sec:AxionMass}.

\begin{figure}[t]
  \begin{center}
    \includegraphics[scale=0.3]{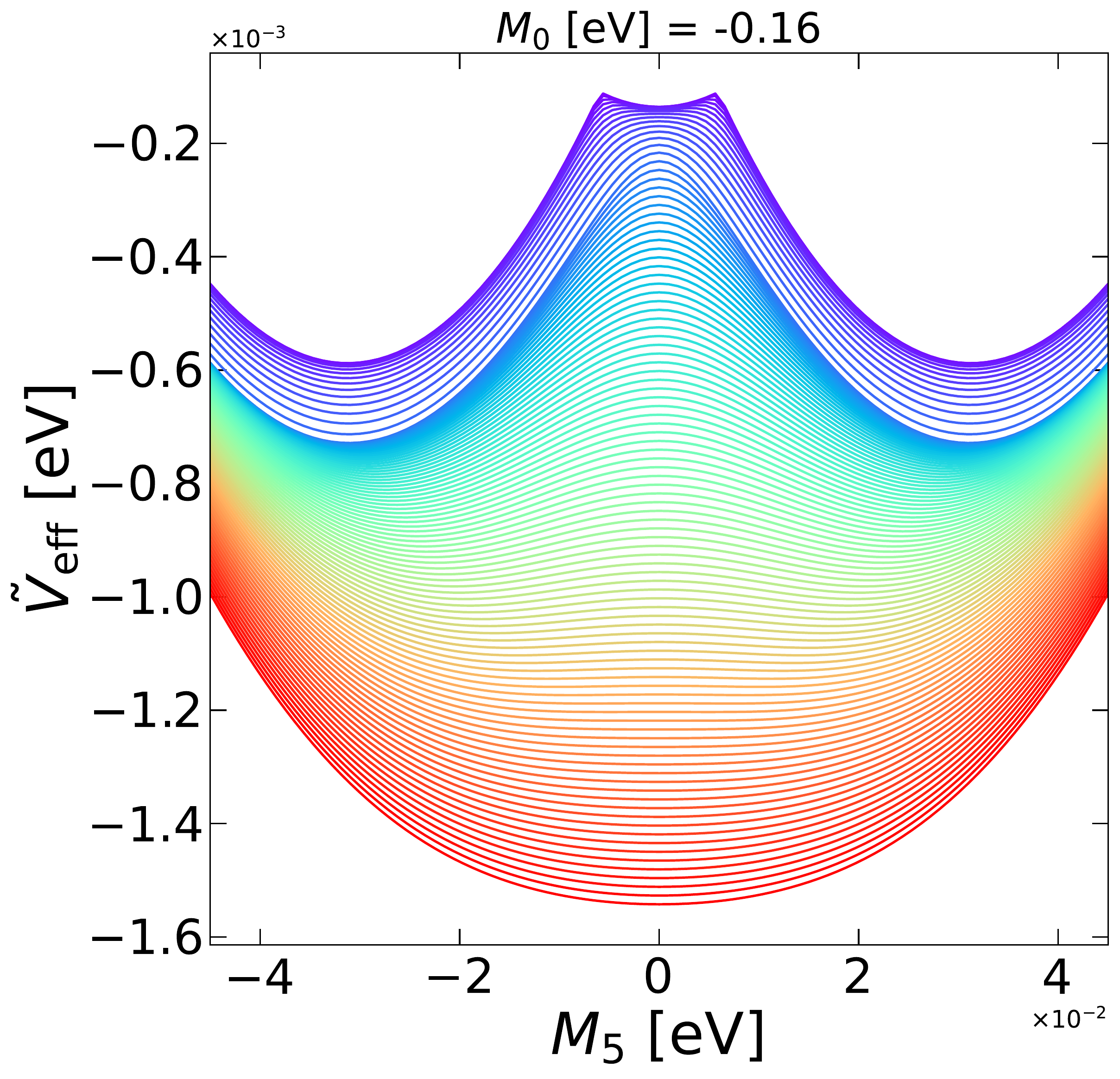}
  \end{center}
  \caption{Normalized effective potential $\tilde{V}_{\rm eff}=[V_{\rm
      eff}(M_f,M_5) -V_{\rm eff}(0,0)]\ell^3$ with $M_f=M_{f0}$ as
    function of $M_5$ for various values of temperature $T=0.1$ to
    100~K from top to bottom. The other parameters are the same as
    Fig.\,\ref{fig:Veff}.}
  \label{fig:Veff_M5}
\end{figure}

\begin{figure}[t]
  \begin{center}
    \includegraphics[scale=0.4]{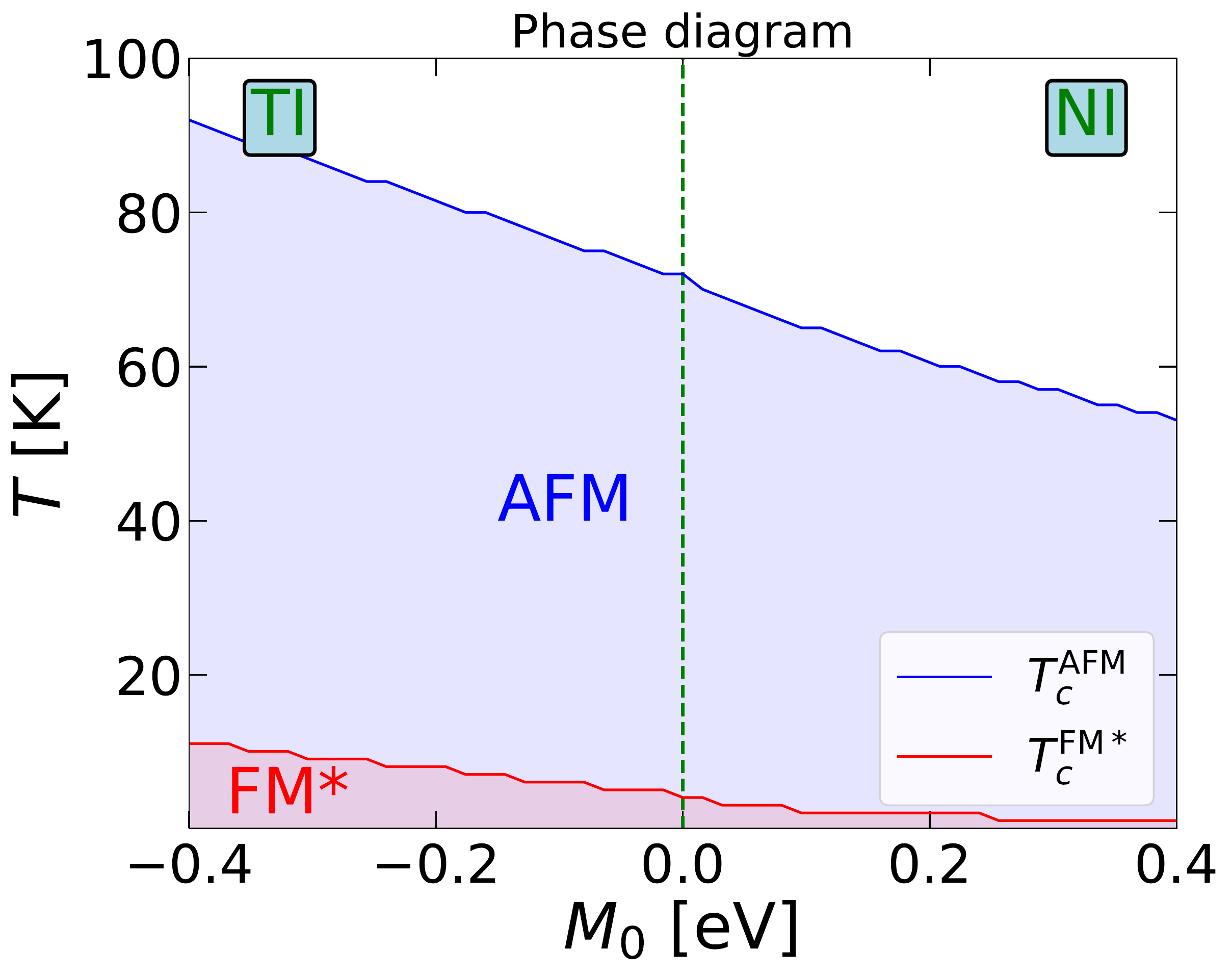}
  \end{center}
  \caption{Phase diagram of the magnetic state. The AFM state and
    metastable FM state are indicated as ``AFM'' and ``FM$*$,''
    respectively.  Here $J^A$, $J^B$, $S$, $x$, $A_i$, and $B_i$
    ($i=1,2$) are taken as the same as Fig.\,\ref{fig:Veff}.  $T^{\rm
      AFM}_c$ is the critical temperature of the AFM state and the
    region $T<T^{\rm AFM}_c$ the AFM state becomes the global
    minimum. $T^{\rm FM*}_c$ is the critical temperature between the
    metastable FM state and the AFM state. As a reference, we
    indicate the topological and normal phases of insulator as ``TI''
    and ``NI,'' respectively, which is separated by a 
    vertical line $M_0=0$ (green dashed).}
  \label{fig:phase}
\end{figure}

Meanwhile the global minimum is the AFM state, we find that at low
temperature there is a local minimum or a metastable point at $M_5=0$
and nonzero $M_f=M_{f0}$, which corresponds to the FM state. Here
$M_{f0}$ is the stationary value for $M_f$ at a given temperature. To
show this explicitly, we compute the effective potential as function
of $M_5$ by taking $M_f=M_{f0}$, which is given in
Fig.\,\ref{fig:Veff_M5}.  In the calculation the other parameters are
the same as Fig.\,\ref{fig:Veff}. The local minimum locates at $M_5=0$
at a sufficiently low temperature and it disappears for $T\gtrsim
10$~K. When the temperature gets even higher, the global minimum
eventually reduces to $(M_f,M_5)=(0,0)$. We note that the result that
the AFM state is the ground state while there is a metastable FM state
is consistent with the first-principles'
calculation for Mn$_2$Bi$_2$Te$_5$~\cite{Li:2020fvr}.

The result that the AFM state is lower than the FM one can be
confirmed analytically as follows. The total energy is calculated from
the free energy,
\begin{align}
  E&=\pdv{\beta}\beta \Omega +\mu Nn
  \nonumber \\
  &=E_e+E_S+H_R\,,
   \label{eq:E_tot}
\end{align}
where $Nn=-\pdv*{\Omega}{\mu}$ and
\begin{align}
  E_e&=\sum_{j,\vb*{k}} (E_{j\vb*{k}}+\mu)n_F(E_{j\vb*{k}})\,,
  \\
  E_S&=-N_sS\left[J^Am^AB_S(S\beta J^Am^A)+J^Bm^BB_S(S\beta J^Bm^B)\right]\,.
\end{align}
Taking the MF values for $M^A$ and $M^B$ at the zero-temperature limit, it
is simply given by
\begin{align}
  E=E_e=-\sum_{\vb*{k}}(e_{1\vb*{k}}+e_{2\vb*{k}})\,.
\end{align}
From Eqs.\,\eqref{eq:e^1} and \eqref{eq:e^2}, it is straightforward to
check that $E|_{M_f=0,M_5=\bar{M}}-E|_{M_f=\bar{M},M_5=0}<0$ for any
values of $\bar{M}$ and $\vb*{k}$.  This is why the AFM is the lowest
energy state. See also Appendix~\ref{sec:ma_zero} for the discussion
of  the FM order for each sublattice.

To get the whole picture, we plot the phase diagram regarding the
magnetic order on $(M_0,T)$ plane in Fig.\,\ref{fig:phase}. The shaded
region shows the AFM or FM states. In the low temperature below
$\order{10^2\,{\rm K}}$, the magnetic state is the AFM. On the other
hand, at sufficiently low temperature that is less than
$\order{10\,{\rm K}}$, the metastable FM state appears. Here we denote
$T_c^{{\rm FM}*}$ as the critical temperature. We note that the stable
AFM state also exists at the temperature, which indicates a possible
phase transition between the metastable FM state and the AFM
state. While the critical temperatures $T^{\rm AFM}_c$ and $T^{\rm
  FM*}_c$ depend on the value of $M_0$, the sign of $M_0$, i.e., the 
topological phase or not, does not have a significant impact on
them. In the next section, we compute the mass of the dynamical axion for
the AFM and the FM states.

\section{Axion mass}
\label{sec:AxionMass}
\setcounter{equation}{0}

As discussed in Ref.\,\cite{Ishiwata:2021qgd}, the dynamical axion
field is defined as the quantum fluctuation around the minimum of the
potential in $M_{5}$ direction. As shown in the previous section there
are two possible minima; the AFM and metastable FM states. Expanding
$M_5$ as $M_5\equiv M_{50}+\varphi$ around the minima, the axion field
$a$ is defined as
\begin{align}
  V_{\rm eff}=V_{\rm eff}(M_{f0},M_{50}) +
  \frac{1}{2}g^2\pdv[2]{V_{\rm eff}}{M_5}\Bigr|_{M_f=M_{f0},M_5=M_{50}}a^2
  +\order{a^4}\,.
\end{align}
Here $g=\dv*{\Phi}{\theta}|_{\theta=\theta_0}$ where $\Phi(\theta)=M_5$ is the
inverse function of $\theta$ defined by~\cite{Li:2009tca}
\begin{align}
   \theta(M_5) = \frac{1}{4\pi} \int d^3k
  \frac{2|d|+d^4}{(|d|+d^4)^2|d|^3}
  \epsilon^{ijkl}d^i
  \partial_{k_x}d^j \partial_{k_y}d^k \partial_{k_z}d^l \,.
  \label{eq:theta_phi}
\end{align}
In the expression we define $|d|^2$ as $|d|^2\equiv
\sum_{a=1}^5d^ad^a$ where $d^5=M_5$ and $\epsilon^{ijkl}$ is the
Levi-Civita symbol with $i,j,k,l$ being 1, 2, 3, and 5. A parameter
$\theta_0$ satisfies $\Phi(\theta_0)=M_{50}$. Then the mass $m_a$ of
the dynamical axion is given by
\begin{align}
  K_a m_a^2=\frac{1}{2V}\pdv[2]{V_{\rm eff}}{M_5}
  \Biggl|_{M_f=M_{f0},M_5=M_{50}}\,,
  \label{eq:Kama2}
\end{align}
where $K_a$ is the stiffness. For reference, see
Appendix~\ref{sec:ma_zero} for the analytic expression of the second
derivative of $V_{\rm eff}$ in the zero-temperature limit. In the
Appendix we check that Eqs.\,\eqref{eq:curvV_Mf} and
\eqref{eq:curvV_M5}, which are the second derivatives with respect to
$M_f$ and $M_5$, respectively, correspond to the spin susceptibility
calculated in the linear perturbation theory.  The stiffness is given
by the perturbative expansion with respect to $\varphi$. Details
are given in Appendix~\ref{sec:stiffness} and the result is
\begin{align}
  K_a=
  \frac{1}{V}\sum_{\vb*{k}}
  \frac{d_0^2}{4(d_0^2+M_{50}^2)^{5/2}}\,,
  \label{eq:KaAFM}
\end{align}
for the AFM states and 
\begin{align}
  K_a =
  \frac{1}{V}\sum_{\vb*{k}}
  \frac{(d_0^2-d_s^2+M_{f0}^2)(e_{2\vb*{k}}-e_{1\vb*{k}})
    +dsM_{f0}(e_{1\vb*{k}}+e_{2\vb*{k}})}
       {8 ds^3M_{f0}e_{1\vb*{k}} e_{2\vb*{k}}}\,,
  \label{eq:KaFM}
\end{align}
for the FM states. It is clear that the functions in the summation are
positive, irrespective of the wavenumber. From Eqs.\,\eqref{eq:Kama2},
\eqref{eq:KaAFM}, and \eqref{eq:KaFM}, we evaluate the axion mass.

\begin{figure}[t]
  \begin{center}
    \includegraphics[scale=0.3]{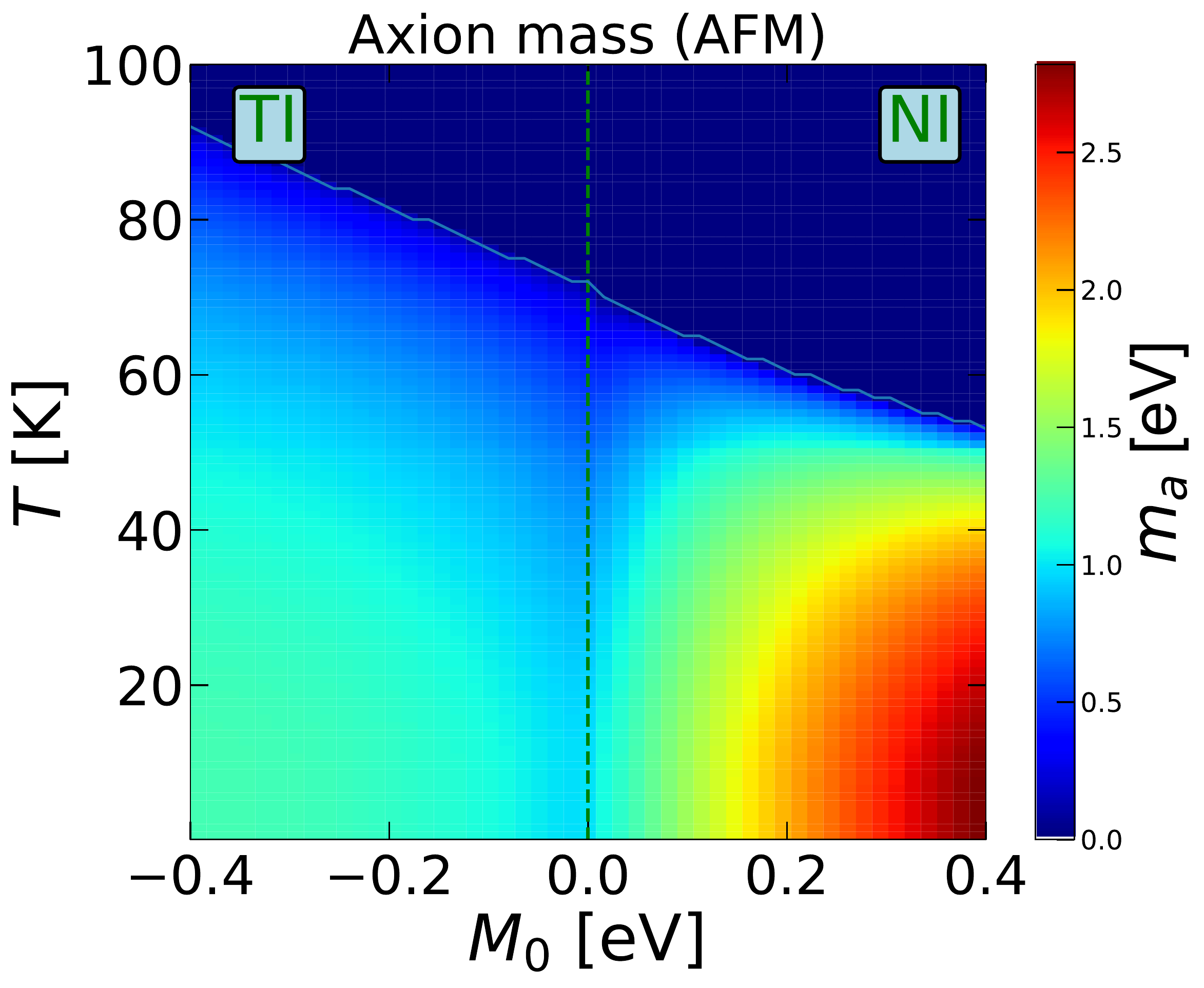}
    \includegraphics[scale=0.3]{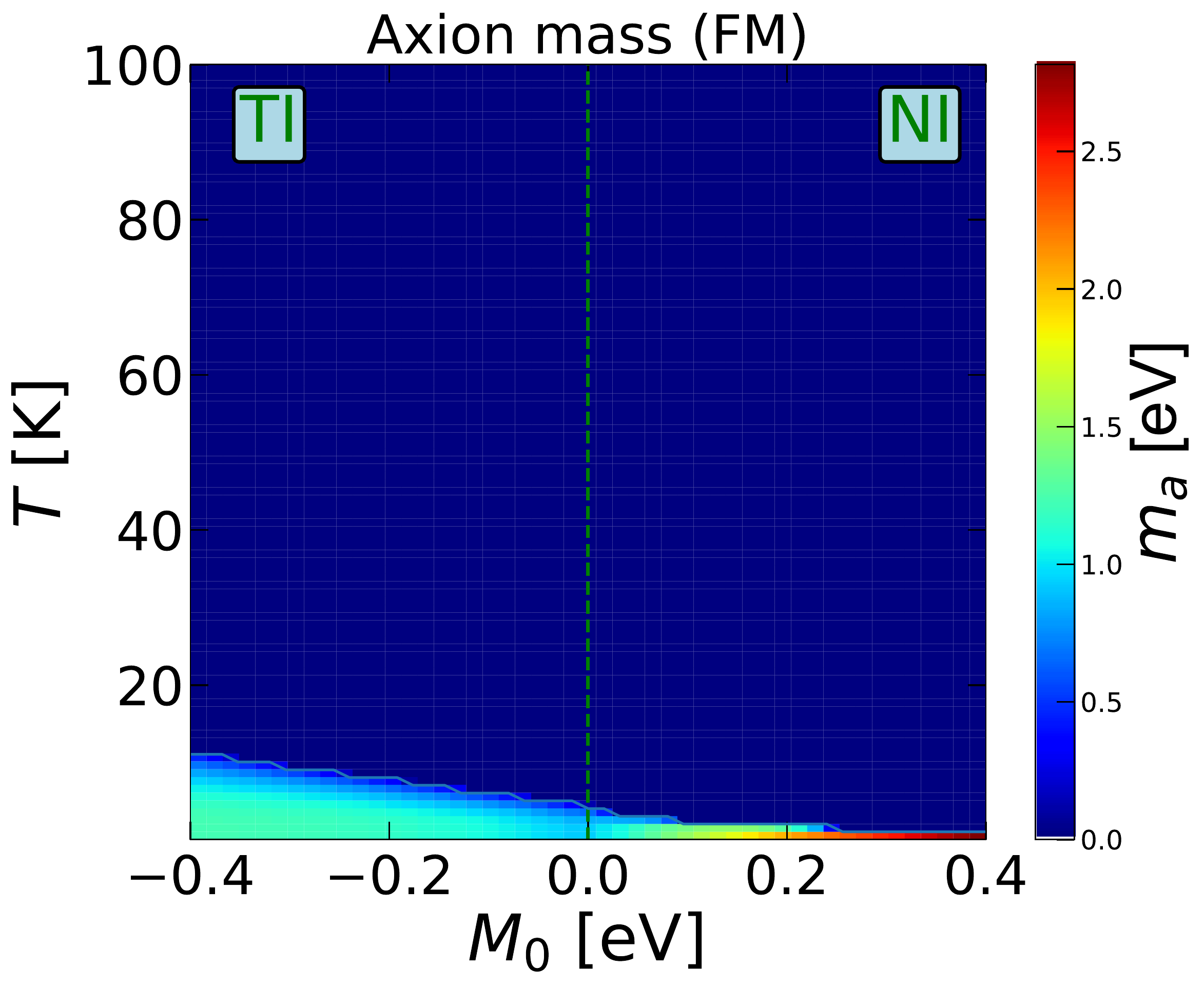}
  \end{center}
  \caption{Color map of the axion mass under the AFM state (left) and
    the metastable FM state (right). Here $J^A$, $J^B$, $S$, $x$,
    $A_i$, and $B_i$ ($i=1,2$) are taken as the same as
    Fig.\,\ref{fig:Veff}. As in Fig.\,\ref{fig:phase}, topological and
    normal phases are indicated as ``TI'' and ``NI,'' respectively. }
  \label{fig:ma}
\end{figure}

As shown in the previous section, there are two possible magnetic
states, the AFM state and metastable FM state. Thus we evaluate the
axion mass for both states.  Fig.\,\ref{fig:ma} shows the axion mass
on the $(M_0,T)$ plane for the AFM state and the metastable FM state. We
find that the axion mass is $\order{{\rm eV}}$ for both states,
except for the phase boundaries. This result is consistent with
Ref.\,\cite{Ishiwata:2021qgd} where only the AFM state is considered
at zero temperature in the Hubbard model.  Here we see a mild
dependence of the mass on the sign of $M_0$. At the phase boundaries,
the axion mass approaches zero from the AFM state to the paramagnetic
state or from the metastable FM state to the AFM state. Therefore the
axion mass can be, in principle, various values by optimizing the
temperature.  This result is quantitatively consistent with
Ref.\,\cite{Wang:2015hhf}, meanwhile the typical value of the axion mass
is different. The typical energy scale of the mass is eV, which is the
same as one estimated in the Hubbard model~\cite{Ishiwata:2021qgd}. On
the other hand, the mass scale is not reduced by the value of the
energy gap, which is a different feature from the results in the
Hubbard model.  The result has a direct impact on the projected mass
range of the {\it particle} axion in the
proposal~\cite{Marsh:2018dlj}. Namely, the targeted mass range is
typically eV, not meV. If the insulator near the phase boundary is
realized, a more suppressed mass range could be probed. However,
preparation of such a state of insulators may not be trivial and there
would be technical challenges for the realistic observation.

It is worth noting that there are two types of axion for the AFM and
FM states. They could be utilized in future {\it particle} axion
search. For example, we speculate that in the circumstance of the FM
state, the {\it particle} axion induces an excitation of the dynamical
axion and it may cause the phase transition to the AFM state, which
can be a possible signal of axion detection. As we mentioned, the
axion mass is not severely influenced by the topology of the
insulators.  This is true for both the AFM and the metastable FM
states. Therefore, various insulators that are not in the topological
phase are also possible candidates of material for the search of the 
{\it particle } axion and axion-like particles. We hope this remark
will inspire future studies for finding realistic materials for {\it
  particle} axion detection experiments.

\section{Conclusion}
\label{sec:conclusion}
\setcounter{equation}{0}

In this study we formulated the mass of the dynamical axion in the
magnetically doped topological insulators.  To this end, we considered
the 3D effective model of TIs with the interaction terms between the
electrons and the impurities.  We found that the antiferromagnetic
state is the ground state at low temperature. Besides, the
ferromagnetic state appears as a metastable state under sufficiently low
temperature.  In both magnetic states, the axion mass is found to be
$\order{\rm eV}$ and it goes to zero as the temperature approaches to
the critical temperature, i.e. the phase boundary. We checked the
results by computing the Van Vleck type spin susceptibility for a band
in linear response theory.  In addition, we found a strong
independence of the energy gap on the axion mass scale. Therefore, the
typical mass scale of the {\it particle} axion search proposed in
Ref.\,\cite{Marsh:2018dlj} should be eV.

The fact that the axion mass can be controlled by temperature may be
suitable for the detection of the {\it particle} axion. Specifically,
the phase boundary has a potential to search the {\it particle} axion
with a suppressed mass. In addition, there can be various magnetic
states indicated in first-principles' calculation, for example, in
Mn$_2$Bi$_2$Te$_5$~\cite{Li:2020fvr}.  To describe such rich magnetic
states rather than the AFM or FM, we need a further extension of the
model, which would be another interesting area of research to find out
the mass of the dynamical axion in a more complicated phase
diagram. We leave it for future work.

\section*{Acknowledgments}

\noindent We thank Makoto Naka for valuable discussions in the early
stage of this project. This work was supported by JSPS KAKENHI Grant
No. JP17K14278, No. JP17H02875, No. JP18H05542, No. JP20H01894, and
the JSPS Core-to-Core Program Grant No. JPJSCCA20200002.

\appendix

\section{Gamma matrices}
\label{sec:Gammas}
\setcounter{equation}{0} 

The Gamma matrices $\Gamma^a$ ($a=1,\cdots, 4$) in Eq.\,\eqref{eq:H0}
are defined as
\begin{align}
  \Gamma^1=\mqty(0& \sigma^1 \\ \sigma^1 & 0)\,,~
  \Gamma^2=\mqty(0& \sigma^2\\ \sigma^2 & 0)\,,~
  \Gamma^3=\mqty(0 & -i \\ i &0)\,,~
  \Gamma^4=\mqty(1 & 0 \\ 0 &-1)\,.
  \label{eq:Gamma1-4}
\end{align}
$\Gamma^5$ is defined by $\Gamma^5=-\Gamma^1\Gamma^2\Gamma^3\Gamma^4$.
In addition we define $\Gamma^{ab}=[\Gamma^a,\Gamma^b]/(2i)$.
To be explicit, they are given by
\begin{align}
  \Gamma^5=\mqty(0&\sigma^3 \\ \sigma^3 &0)\,,~
  \Gamma^{12}=\mqty(\sigma^3 &0 \\ 0 &\sigma^3)\,.
\end{align}

In the sublattice basis, the Gamma matrices are given by
\begin{align}
  \Gamma^{1\,\prime}=\mqty(\sigma^1 & 0 \\ 0 & -\sigma^1)\,,~
  &\Gamma^{2\,\prime}=\mqty(\sigma^2 & 0 \\ 0 & -\sigma^2)\,,~
  \Gamma^{3\,\prime}=\mqty(0 & -i \\ i &0)\,,~
  \Gamma^{4\,\prime}=\mqty(0 & -1 \\ -1 &0)\,, \\
 & \Gamma^{5\,\prime}=\mqty(\sigma^3 &0 \\ 0 &-\sigma^3)\,,~
  \Gamma^{12\,\prime}=\mqty(\sigma^3 &0 \\ 0 &\sigma^3)\,.
\end{align}

\section{Effective action}
\label{sec:Effaction}
\setcounter{equation}{0} 

Since we consider the half-filling case, i.e., the number of electrons
is fixed, the grand potential $\Omega_e (M^A,M^B)$ of the
electron part corresponds to the Helmholtz free energy
$F_e(M^A,M^B)$. The Gibbs free energy $G_e(m^A,m^B)$ is then
given by the Legendre transformation:
\begin{align}
  G_e(m^A,m^B) &= -\sum_{I=A,B} \pdv{F_e}{M^I} M^I + F_e
  \nonumber \\
  &= -N\sum_{I=A,B}J^Im^I M^I + F_e
  \,,
\end{align}
where $m^I\equiv (1/N)\pdv*{F_e}{J^I M^I}$ ($I=A,B$).  This definition
is equivalent to the MF values for $m^A$ and $m^B$ given in
Eqs.\,\eqref{eq:mAMF} and \eqref{eq:mBMF}. In addition $G_e(m^A,m^B)$
is equivalent to $\Omega_e+H_R$ where $m^I$ are taken to be their the
MF values. Since the Gibbs free energy corresponds to the effective
action in the quantum field theory, Eq.\,\eqref{eq:Seff} is considered
as the effective action. In our analysis we give Eq.\,\eqref{eq:Seff}
as function of $M_f$ and $M_5$, i.e., $m^I=m^I(M^I)$, instead of $m^I$
itself since we are interested in the dynamical field around the
minimum $(M_{f0},M_{50})$.

\section{The mass at zero temperature and spin susceptibility}
\label{sec:ma_zero}
\setcounter{equation}{0} 

At the zero temperature, $\Omega_S+H_R$ in $\Omega$ cancels. Then the
curvature around the minimum is computed easily. The results are
\begin{align}
  \frac{1}{2}\pdv[2]{\Omega}{M_f}\Bigl|_{M_f=M_{f0},M_5=M_{50}}
  &=
  \frac{1}{2}\pdv[2]{\Omega_e}{M_f}\Bigl|_{M_f=M_{f0},M_5=M_{50}}
  \nonumber \\
  &=
  \frac{1}{2}\sum_{\vb*{k}}(d_0^2-d_s^2)
  \left[\frac{1}{e_{1\vb*{k}}^3}+\frac{1}{e_{2\vb*{k}}^3}
    \right]_{M_f=M_{f0},M_5=M_{50}} \,.
  \label{eq:curvV_Mf}
  \\
  \frac{1}{2}\pdv[2]{\Omega}{M_5}\Bigl|_{M_f=M_{f0},M_5=M_{50}}
  &=
  \frac{1}{2}\pdv[2]{\Omega_e}{M_5}\Bigl|_{M_f=M_{f0},M_5=M_{50}}
  \nonumber \\
  &= \frac{1}{2}\sum_{\vb*{k}}
  \Biggl[
    -M_5^2\Biggl\{
    \frac{\Bigl(1+\frac{M_f}{\sqrt{d_s^2+M_5^2}}\Bigr)^2}{e_{1\vb*{k}}^3}+
    \frac{\Bigl(1-\frac{M_f}{\sqrt{d_s^2+M_5^2}}\Bigr)^2}{e_{2\vb*{k}}^3}
    \Biggr\}
    \nonumber \\ &~~~+
     \frac{1+\frac{M_fd_s^2}{(d_s^2+M_5^2)^{3/2}}}{e_{1\vb*{k}}}
    +\frac{1-\frac{M_fd_s^2}{(d_s^2+M_5^2)^{3/2}}}{e_{2\vb*{k}}}
    \Biggr]_{M_f=M_{f0},M_5=M_{50}} \,.
  \label{eq:curvV_M5}
\end{align}
It is clear that both quantities are positive and we checked that
$\pdv*{\Omega}{M_f}{M_5}= 0$ at the minimum. Therefore, the
minimum is stable.

The above results can be confirmed by the spin susceptibility of
electron computed in linear response theory. In the magnetic TIs, the
local spins have an effective interaction via electrons.  In the
present case it corresponds to $-\tilde{J}^{\rm eff}_f M_fM_f$ and
$-\tilde{J}^{\rm eff}_5 M_5M_5$, where the normalized effective exchange
couplings are given by $\tilde{J}^{\rm eff}_f=\chi^e_f/2$ and
$\tilde{J}^{\rm eff}_5=\chi^e_5/2$~\cite{Wang:2014}.\footnote{A factor
of 2 is different compared to the expression given in
Ref.\,\cite{Wang:2014}. We start with the Hamiltonian
$1/(2\chi^e_{f})m_t^2+1/(2\chi^e_{5})m_r^2-(M_fm_t+M_5m_r)$ to obtain
$-(1/2)\chi^e_fM_fM_f-(1/2)\chi^e_5M_5M_5$.} Here $\chi^e_f$ and
$\chi^e_5$ are the Van Vleck-type spin susceptibility for a band
insulator. Namely $\chi^e_{f}$ and $\chi^e_{5}$ correspond to the
squared mass parameters. By taking ${\cal H}^m_{\vb*{k}}$ as the 
perturbation in liner response theory they are calculated as
\begin{align}
  \chi^e_f &= \sum_{\vb*{k},m,n}[n_F(E_{n\vb*{k}})-n_F(E_{m\vb*{k}})]
  \frac{\bra{u_{n\vb*{k}}}\Gamma^{12}\ket{u_{m\vb*{k}}}
    \bra{u_{m\vb*{k}}}\Gamma^{12}\ket{u_{n\vb*{k}}}}
       {E_{m\vb*{k}}-E_{n\vb*{k}}}\,,
       \label{eq:chi_f}
       \\
  \chi^e_5 &= \sum_{\vb*{k},m,n}[n_F(E_{n\vb*{k}})-n_F(E_{m\vb*{k}})]
  \frac{\bra{u_{n\vb*{k}}}\Gamma^{5}\ket{u_{m\vb*{k}}}
    \bra{u_{m\vb*{k}}}\Gamma^{5}\ket{u_{n\vb*{k}}}}
       {E_{m\vb*{k}}-E_{n\vb*{k}}}\,,
       \label{eq:chi_5}
\end{align}
where $E_{n\vb*{k}}$ and $\ket{u_{n\vb*{k}}}$ are the energy
eigenvalues and eigenstates of the electron, respectively, which is
obtained by diagonalizing ${\cal H}_{\vb*{k}}^{\rm TI}$.

To compare the spin susceptibilities given in Eqs.\,\eqref{eq:chi_f}
and \eqref{eq:chi_5}, it is appropriate to start with the free energy
$\Omega'\equiv \Omega_e-N(M_fm_t+M_5m_r)$, where the free energy for
the local spin is omitted. This is because the spin susceptibilities
given above are obtained in the linear response theory by taking the
$-(M_fm_t+M_5m_r)$ term as the perturbation while $H_e$ is the zero-th
order Hamiltonian. Then, it is straightforward to get
\begin{align}
  \frac{1}{2}\pdv[2]{\Omega'}{M_f}\Bigl|_{M_f=M_{f0},M_5=M_{50}}
  &=
  -\frac{1}{2}\pdv[2]{\Omega_e}{M_f}\Bigl|_{M_f=M_{f0},M_5=M_{50}}\,,
  \\
  \frac{1}{2}\pdv[2]{\Omega'}{M_5}\Bigl|_{M_f=M_{f0},M_5=M_{50}}
  &=
  -\frac{1}{2}\pdv[2]{\Omega_e}{M_5}\Bigl|_{M_f=M_{f0},M_5=M_{50}}\,,
\end{align}
In fact we confirmed that
\begin{align}
  &\frac{1}{2}\pdv[2]{\Omega_e}{M_f}\Bigl|_{M_f=0,M_5=0}
  =\frac{1}{2}\chi^e_f=\sum_{\vb*{k}}\frac{1-d_s^2/d_0^2}{d_0}\,,
  \\
  &\frac{1}{2}\pdv[2]{\Omega_e}{M_5}\Bigl|_{M_f=0,M_5=0}
   =\frac{1}{2}\chi^e_5=\sum_{\vb*{k}}\frac{1}{d_0}\,.
\end{align}
Moreover, since Eqs.\,\eqref{eq:chi_f} and \eqref{eq:chi_5} from the
linear respose theory can be applied for a generic form of the
Hamiltonian, it would be possible to take ${\cal
  H}_{e\vb*{k}}|_{M_f=M_{f0},M_5=M_{50}}$ and $\delta
M_f\Gamma^{12}+\delta M_5\Gamma^5$ as the primary Hamiltonian and
perturbation, respectively. See also Appendix~\ref{sec:stiffness} for
such an expansion. In that case, $E_{n\vb*{k}}$ and
$\ket{u_{n\vb*{k}}}$ correspond to the energy eigenvalues and
eigenstates of the Hamiltonian ${\cal H}_{e\vb*{k}}$ where
$M_f=M_{f0}$ and $M_5=M_{50}$. By computing $\chi^e_f$ and $\chi^e_5$
numerically, we checked the correspondence between the mass squared
and the Van Vleck-type spin susceptibility,
\begin{align}
  &\frac{1}{2}\pdv[2]{\Omega_e}{M_f}\Bigl|_{M_f=M_{f0},M_5=M_{50}}
  =\frac{1}{2}\chi^e_f\,,
  \\
  &\frac{1}{2}\pdv[2]{\Omega_e}{M_5}\Bigl|_{M_f=M_{f0},M_5=M_{50}}
   =\frac{1}{2}\chi^e_5\,,
\end{align}
for any value of $M_{f0}$ and $M_{50}$.

Additionally, we checked that the spin susceptibilities for
the sublattice $A$ and $B$, which are defined by
\begin{align}
  \chi^e_A &= \sum_{\vb*{k},m,n}[n_F(E_{n\vb*{k}})-n_F(E_{m\vb*{k}})]
  \frac{\bra{u_{n\vb*{k}}}(\Gamma^{12}+\Gamma^5)/2\ket{u_{m\vb*{k}}}
    \bra{u_{m\vb*{k}}}(\Gamma^{12}+\Gamma^5)/2\ket{u_{n\vb*{k}}}}
       {E_{m\vb*{k}}-E_{n\vb*{k}}}\,,
       \label{eq:chi_A}
       \\
  \chi^e_B &= \sum_{\vb*{k},m,n}[n_F(E_{n\vb*{k}})-n_F(E_{m\vb*{k}})]
  \frac{\bra{u_{n\vb*{k}}}(\Gamma^{12}-\Gamma^5)/2\ket{u_{m\vb*{k}}}
    \bra{u_{m\vb*{k}}}(\Gamma^{12}-\Gamma^5)/2\ket{u_{n\vb*{k}}}}
       {E_{m\vb*{k}}-E_{n\vb*{k}}}\,,
       \label{eq:chi_B}
\end{align}
are both positive. This result is also expected since the order of
each sublattice is the FM.

\section{Propagator and the stiffness}
\label{sec:stiffness}
\setcounter{equation}{0} 

The stiffness is given by the coefficient of the axion kinetic
term.\footnote{Ref.\,\cite{Shiozaki:2013wda} gives a similar
calculation using the Hubbard-Stratonovich transformation but to discuss
topological superconductors and superfluids. See also
Ref.\,\cite{Roy:2015xua} for the renormalization group approach. } To
give the kinetic term we consider a fluctuation of $M_5$ around the
stationary point by promoting $M_5$ as a dynamical degree of freedom.
To make the discussion generic, we take $M_{5}=M_{50}+\varphi$. The
kinetic term is obtained by expanding ${\cal S}_e$ with respect to
$\varphi$. To this end, we write ${\cal H}_e={\cal H} + \delta {\cal
  H}$. In the wavenumber space, they are defined as
\begin{align}
  {\cal H}_{\vb*{k}} &= {\cal H}_{\vb*{k}}^{\rm TI} + M_{f0}
  \Gamma^{12} + M_{50} \Gamma^{5}\,,
  \\
  \delta {\cal H}_{\vb*{k}} &= \varphi \Gamma^5\,.
\end{align}
Using $\ln \det [\partial_\tau+{\cal H}_e]=\Tr \ln
[\partial_\tau+{\cal H}_e]$ and 
\begin{align}
  {\rm Tr}\ln(\partial_\tau+{\cal H}+\delta {\cal H})
  ={\rm Tr}\ln (-G^{-1})-\sum_{n=1}^\infty {\rm Tr}(G\delta {\cal H})^n\,,
  \label{eq:Tr_expand}
\end{align}
where $G^{-1}=-\partial_\tau-{\cal H}$, the kinetic term is obtained
from the quadratic term in the second term of
Eq.\,\eqref{eq:Tr_expand}:
\begin{align}
  {\rm Tr}(G \delta {\cal H})^2 =
  \frac{V^2}{N^2}
  \int^\beta_0 d\tau_i\int^\beta_0 d\tau_j \sum_{x_i,x_j}
  {\rm Tr}[G(x_i-x_j)\delta{\cal H}(x_j)G(x_j-x_i)\delta{\cal H}(x_i)]\,.
\end{align}
Here the arguments of $G$ and $\delta{\cal H}$ represent
$x_i=(\tau_i,\vb*{x}_i)$. The propagator and the field $\delta{\cal
  H}$ are expanded as
\begin{align}
  G(x_i-x_j)&=\frac{1}{\beta V}\sum_{i\omega_{n}}
  \sum_{\vb*{k}}\tilde{G} (k)
  e^{-i\omega_{n}(\tau_1-\tau_2)+i\vb*{k}\vdot (\vb*{x}_i-\vb*{x}_j)}\,,
  \\
   \delta {\cal H}(x_i)&=\frac{1}{\beta V}\sum_{i\omega_{n}}
   \sum_{\vb*{k}}\tilde{\delta {\cal H}} (k)
  e^{-i\omega_{n}\tau+i\vb*{k}\vdot \vb*{x_i}}\,,
\end{align}
where $\tilde{G}^{-1}(k)=i\omega_n-{\cal H}_{\vb*{k}}$. Similarly
to $x_i$, we take the argument $k_i$ of $\tilde{G}$ and $\tilde{\delta
  {\cal H}}$ as $k_i=(i\omega_{ni},\vb*{k}_i)$. Then
\begin{align}
  {\rm Tr}(G \delta {\cal H})^2 =
  \frac{1}{\beta^2 V^2}\sum_{i\omega_{n1}}\sum_{i\omega_{n2}}
  \sum_{\vb*{k}_1,\vb*{k}_2}
       {\rm Tr}[\tilde{G}(k_1)\tilde{\delta {\cal H}}(k_2)
         \tilde{G}(k_1-k_2)\tilde{\delta {\cal H}}(-k_2)]\,.
\end{align}

We find the propagator $\tilde{G}$ in the momentum space is given by
\begin{align}
  \tilde{G}(i\omega_n,\vb*{q})=\frac{1}{F}\left[(i\omega_n -\epsilon)g_0
  +\sum_{a=1}^5g_1^a \Gamma^a+\sum_{a=1}^4g_2^a \Gamma^a\Gamma^5
  +\sum_{ab}g^{ab}\Gamma^{ab}\right]\,,
\end{align}
where $\epsilon=\epsilon_0-\mu$ and 
\begin{align}
  g_0 &= (i\omega_n-\epsilon)^2-d_0^2-M_{50}^2-M_{f0}^2\,,
  \\
  g_1^a&=-d^a\{ -(i\omega_n-\epsilon)^2 +d_0^2+M_{50}^2+M_{f0}^2\}~~~(a=1,2)\,,
  \\
  g_1^a&=-d^a\{ -(i\omega_n-\epsilon)^2 +d_0^2+M_{50}^2-M_{f0}^2\}~~~(a=3,4,5)\,,
  \\
  g_2^1&=2id^2M_{50}M_{f0}\,,
  \\
  g_2^2&=-2id^1M_{50}M_{f0}\,,
  \\
  g_2^3&=2i(i\omega_n-\epsilon)d^4M_{f0}\,,
  \\
  g_2^4&=-2i(i\omega_n-\epsilon)d^3M_{f0}\,,
  \\
  g^{12}&=\{(i\omega_n-\epsilon)^2-(d^1)^2-(d^2)^2+(d^3)^2+(d^4)^2
  +M_{50}^2-M_{f0}^2\}M_{f0}\,,
  \\
  g^{34}&=2(i\omega_n-\epsilon)M_{50}M_{f0}\,,
  \\
  g^{23}&=2d^1d^3M_{f0}\,,
  \\
  g^{13}&=-2d^2d^3M_{f0}\,,
  \\
  g^{14}&=-2d^2d^4M_{f0}\,,
  \\
  g^{24}&=2d^1d^4M_{f0}\,,
  \\
  F&=\{(i\omega_n-\epsilon)^2-|d_0|^2-(M_{f0}-M_{50})^2\}
  \{(i\omega_n-\epsilon)^2-|d_0|^2-(M_{f0}+M_{50})^2\}
  -4d_s^2M_{f0}^2\,.
\end{align}
Using the propagator, the stiffness is given by
\begin{align}
  K_a=-\frac{1}{2\beta V}
  \pdv[2]{(i \omega_{nk})}\,
  \sum_{i\omega_{nq},\vb*{q}}{\rm Tr}[\tilde{G}(i\omega_{nq},\vb*{q})
    \Gamma^5
    \tilde{G}(i\omega_{nq}+i\omega_{nk},\vb*{q})\Gamma^5]
  \Bigr|_{i\omega_{nk}=0}\,.
  \label{eq:Tr2_forK}
\end{align}
Here we redefined $k_1$ and $k_2$ as $k_1=q$ and $k_2=-k$ and
taken $\vb*{k}=0$ in $\tilde{G}(q+k)$. This expression can be checked
by taking $M_{f0}=0$ to get
\begin{align}
  K_a=
  \frac{1}{V}\sum_{\vb*{q}}
  \frac{d_0^2}{4(|d_0|^2+M_{50}^2)^{5/2}}\,,
  \label{eq:Ka_AFM}
\end{align}
in the zero-temperature limit, which agrees with the one given in
Ref.\,\cite{Ishiwata:2021qgd}. This is the expression of the stiffness
for the AFM state. The FM state corresponds to nonzero $M_{f0}$ and
$M_{50}=0$. In that case, we find in the zero-temperature limit
\begin{align}
  K_a =
  \frac{1}{V}\sum_{\vb*{q}}
  \frac{(d_0^2-d_s^2+M_{f0}^2)(e_{2\vb*{q}}-e_{1\vb*{q}})
    +dsM_{f0}(e_{1\vb*{q}}+e_{2\vb*{q}})}
       {8 ds^3M_{f0}e_{1\vb*{q}} e_{2\vb*{q}}}\,,
  \label{eq:Ka_FM}
\end{align}
where $M_5=0$ is taken in $e_{1\vb*{q}}$ and $e_{2\vb*{q}}$. We note
that the zero-temperature limit is a good approximation since we
discuss the system at up to $\order{10^2}\,{\rm K}$, which is smaller
than the typical energy scale of the electron energy.

The formalism given in the discrete space can be written in the
continuum case by the following replacements:
\begin{align}
  \frac{1}{N}\sum_i\to \frac{1}{V}\int d^3x \,,~~~
 \frac{1}{V}\sum_{\vb*{k}} \to \int \frac{d^3k}{(2\pi)^3}\,,
  \label{eq:sum_to_integ} \\
  \sqrt{N} c_i \to \sqrt{V}\psi(\vb*{x})\,,~~~
  N n_{i} \to V n(\vb*{x})\,.
  \label{eq:c_to_psi}
\end{align}
Here $\psi$ is the wavefunction of the electrons in the continuum space,
$n=\psi^\dagger \psi$, and $n_i=c_i^\dagger c_i$.

\end{document}